\documentclass[twocolumn,showpacs, preprintnumbers, amsmath, amssymb, superscriptaddress,pra]{revtex4}
\usepackage{graphicx}
\usepackage{dcolumn}
\usepackage{bm}
\usepackage{epsf}
\usepackage{amssymb}
\DeclareGraphicsExtensions{.eps}
\usepackage{epstopdf}
\usepackage{natbib}
\usepackage{subfigure}
\usepackage{graphicx}
\begin{document}

\author{David S. Simon}
\affiliation{Dept. of Electrical and Computer Engineering, Boston
University, 8 Saint Mary's St., Boston, MA 02215}
\affiliation{Dept. of Physics and Astronomy, Stonehill College, 320 Washington Street, Easton, MA 02357}
\author{Casey Fitzpatrick}
\affiliation{Dept. of Electrical and Computer Engineering, Boston
University, 8 Saint Mary's St., Boston, MA 02215}
\author{Alexander V. Sergienko}
\affiliation{Dept. of Electrical and Computer Engineering, Boston
University, 8 Saint Mary's St., Boston, MA 02215}
\affiliation{Photonics Center, Boston
University, 8 Saint Mary's St., Boston, MA 02215}
\affiliation{Dept. of Physics, Boston University, 590 Commonwealth
Ave., Boston, MA 02215}

\begin{abstract}
Security against simple eavesdropping attacks is demonstrated for a recently proposed quantum key distribution protocol which uses the Fibonacci
recursion relation to enable high-capacity key generation with entangled photon pairs. No transmitted pairs need to be discarded in reconciliation; the only pairs not used for key generation are those used for security-checking. Although the proposed approach does not allow eavesdropper-induced errors
to be detected on single trials, it can nevertheless reveal the eavesdropper's action on the quantum channel by detecting changes in the distribution of outcome probabilities over multiple trials, and can do so as well as the BB84 protocol. The mutual information shared by the participants is calculated and used to show that a secret key can always be distilled.
\end{abstract}

\title{Security in the Multi-Dimensional Fibonacci Protocol}


\maketitle


\section{Introduction}\label{introsection}

In quantum key distribution (QKD), two legitimate users of the system, Alice and Bob, attempt to generate a shared encryption key in such a way that the laws of
quantum mechanics prevent an unauthorized eavesdropper, Eve, from obtaining the key without revealing her presence. Restricting ourselves here to systems that
use pairs of entangled photons, the most common approach is the Ekert protocol \cite{ekert}, which is itself an entangled version of the earlier BB84 protocol
\cite{bb84}. An embodiment of this approach using photon polarization works as follows. Alice and Bob each receive half of the entangled pair from a common
source. Often the source is in Alice's lab, but it may be under the control of a third party (or even under the control of Eve herself). As the photons arrive,
Alice and Bob each randomly choose one of two bases in which to make a linear polarization measurement. One basis consists of horizontal and diagonal
polarization states ($|H\rangle$ and $|V\rangle$), the other involves diagonal states ($\nearrow\rangle$) and ($|\searrow \rangle$) polarized at $45^\circ$ from
the horizontal and vertical. They then communicate on a classical (possibly public) channel in order to compare their bases (but not the results of their
measurements), discarding those trials in which they used different bases.

In the simplest possible eavesdropping attack, Eve also makes a polarization measurement of one photon as it is en route to Bob. She must guess randomly which
measurement basis to use. If she guesses correctly, measuring in the same basis as Alice and Bob, then she gains full information about the polarization state
and therefore has complete information about the state of the corresponding key segment. Half of the time she guesses incorrectly, in which case her outcome is
completely random and uncorrelated with Alice's result. When she sends on the photon, Bob's results will also be completely randomized in the original basis.
This disturbance in Bob's results allows Eve's actions to be revealed.  Alice and Bob randomly select a subset of their results for a security check, exchanging
the results of their measurements as well as the basis used on these trials. If Eve intercepts a fraction $\eta$ of Bob's photons she introduces an error rate of
${\eta\over 4}$ between Alice's outcomes and Bob's in the security-checking subset.

The polarization states have an effective Hilbert space of dimension 2, allowing a single bit of the key to be extracted in each polarization measurement. There
have been a number of attempts to generalize the procedure above with other physical degrees of freedom that have higher-dimensional effective Hilbert spaces,
thereby allowing more than one bit of the key to be generated per photon \cite{bruss,bechmann,bour,cerf,groblacher}, as well as improved ability to detect
eavesdroppers.  One particularly promising way \cite{mair,vaziri,simonJswitching} to achieve this goal is to use the photon's orbital angular momentum (OAM)
\cite{francke,allen,molina} instead of polarization. The OAM about the propagation axis is quantized, $L_z=l\hbar$, where the topological charge $l$ can take on
any integer value. If a range of OAM values of size $N$ is used as the alphabet (for example $l=1,2,\dots ,N$ or $l=-{{N-1}\over 2},\dots , +{{N-1}\over 2}$ for
even $N$), then each photon can be used to determine up to $log_2N$ bits of the key.  Although in principle such schemes allow unlimited numbers of bits per
photon, their experimental complexity increases rapidly with increasing $N$. For the most part we will work in this paper with arbitrary $N$, but in order to
make the discussion more concrete we will at some points restrict ourselves to an alphabet of size $N=8$ (capable of achieving $\log_28 =3$ bits per photon). The
generalization to higher values of $N$ is straightforward.

In \cite{simon1}, an approach to high-dimensional QKD was proposed based on a novel entangled light source
\cite{liew,trevino,trevino2,dalnegro,lawrence} that produces output with absolute values of the OAM spectrum restricted to the Fibonacci sequence. (Recall that
the Fibonacci sequence starts with initial values $F_1=1$ and $F_2=2$, generating the rest of the sequence via the recurrence relation $F_{n+2}=F_{n+1}+F_n$.)
These states pump a nonlinear crystal, leading to spontaneous parametric down conversion (SPDC), in which a small proportion of the input
photons are split into entangled signal and idler output photons. After the crystal, any photons with non-Fibonacci values of OAM are filtered
out. Angular momentum conservation and the Fibonacci recurrence relation conspire to force the two output photons to have OAM values that are adjacent
Fibonacci numbers (for example $F_m$ and $F_{m+1}$). These photons can then be used to generate a secret key, as detailed in the next section. 

The Fibonacci protocol requires the ability to distinguish between single OAM eigenstates and pairwise superpositions of eigenstates. This cannot be done
unambiguously on a single trial; however, as shown in \cite{simon2}, an interferometric approach allows statistical discrimination of the possibilities over
multiple trials. Using this approach, we show in this paper that the outcome probability distributions can be built up in such a way that eavesdropping alters
them in a detectable manner. In this way, an eavesdropper can be revealed over multiple trials even though it may not be possible to identify errors on any
individual trial. This is especially important because even without eavesdropping there is some probability of error in discriminating between states in
individual trials, due to the non-orthogonality of the relevant superposition states. This complication must be dealt with in the reconciliation stage.    BB84
and most other protocols rely on the rate of errors in individual outcomes for security, but here we must rely on statistical changes wrought by the
eavesdropper's actions over many trials. As a result, rather than using eavesdropper-induced error rates as a measure of eavesdropper detection, it is more
appropriate here to use probability-based measures such as the disturbance ${\cal D}$ \cite{lutken96}, which is defined in section \ref{probsection}.

Section \ref{setupsection} reviews the procedure for implementing the Fibonacci protocol in the context of OAM measurements.   The effect of eavesdropping is examined in section \ref{evesection}. Details of the means of dealing
with the superposition states are described in section \ref{supersection}, with calculations
of various measures of security carried out in section \ref{probsection} and conclusions in Section \ref{conclusionsection}. A brief discussion of key
reconciliation over a classical channel is given in Appendix A, and detailed expressions for the probability distributions of outcomes for the protocol appear in
Appendix B.

\section{Setup for Key Distribution}\label{setupsection}

Here, we review the Fibonacci protocol \cite{simon1}. The protocol discussed here is a slightly improved variation on the original proposal of \cite{simon1}, making use of the apparatus described in more detail in \cite{simon2}. Note that no trials need to be
discarded in the scheme described, aside from those used for security checking. This is in contrast to the original version of the Fibonacci protocol
\cite{simon1} where $1\over 4$ of the trials were discarded during reconciliation and the BB84/Ekert protocol, where ${1\over 2}$ must be discarded.

The Fibonacci protocol allows multiple key bits to be generated per photon. The required basis modulation can be done in a completely passive manner, and
switching only has to be done between two fixed bases, so that the complexity of the setup increases much more slowly with increasing $N$ than in other OAM-based
QKD methods, such as \cite{groblacher}.

\begin{figure}
\begin{center}
\includegraphics[scale=.28]{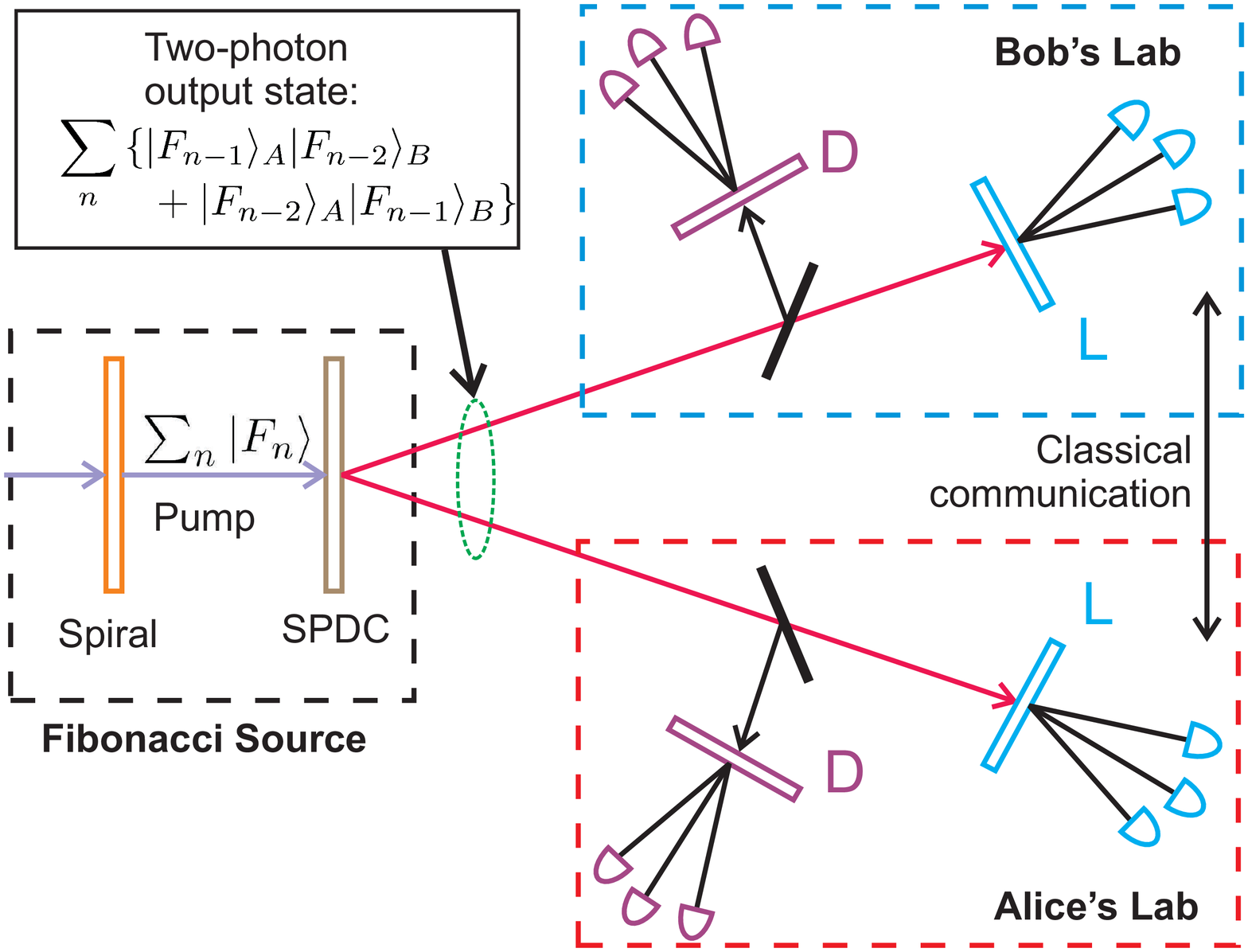}
\caption{Schematic setup for generating cryptographic key with Fibonacci-valued OAM states.}
\label{setupfig}
\end{center}
\end{figure}

The basic setup is shown schematically in Fig. \ref{setupfig}. The source on the left (see \cite{simon1} for more detail) uses a down conversion source and
scattering from an aperiodic spiral nanoarray \cite{liew,trevino,trevino2,dalnegro,lawrence} to create an entangled superposition of states with Fibonacci-valued OAM states in the two output directions:
\begin{equation}\psi =\sum_n \big( |F_{n-1}\rangle_A |F_{n-2}\rangle_B + |F_{n-2}\rangle_A |F_{n-1}\rangle_B\big) ,\label{initialstate}\end{equation} where the
index $n$ runs over the indices of the allowed Fibonacci numbers in the pump beam: $|\Psi\rangle_{pump} =\sum_n |F_n\rangle.$  Alice and Bob each receive one
photon from the entangled pair. In each of their labs is a  50/50 beam splitter which randomly directs the photon to one of two types of detection stages. The
stage labeled ``$\mathbb{L}$-type'' detection consists of an OAM sorter \cite{leach,berkhout,lavery} followed by a set of single-photon detectors. The OAM sorter
sends OAM eigenstates with different $l$ values into different outgoing directions, so that they will be registered in different detectors, thus allowing $l$ to
be determined. The other type of detection (labeled ``$\mathbb{D}$ type'') is used to distinguish different superpositions $|S_n\rangle $ of the form
\begin{eqnarray}|S_n\rangle ={1\over \sqrt{2}}\big( |F_{n-1}\rangle+|F_{n+1}\rangle \big) .\end{eqnarray} The detection of
such superposition states can be accomplished in several ways \cite{miyamoto,jack,kawase}. One complication with the $\mathbb{D}$-type detection is that adjacent
superposition states, such as ${1\over \sqrt{2}}\left( |F_{n-1}\rangle+|F_{n+1}\rangle \right) $ and ${1\over \sqrt{2}}\left( |F_{n+1}\rangle+|F_{n+3}\rangle
\right) $ are not orthogonal, meaning that the two states cannot be unambiguously distinguished from each other. This adds complications to the classical exchange and alters the corresponding detection probabilities; however we will
see below that this extra complication is well compensated by increased the key-generating capacity. Moreover, the degree of complication does not grow with the
size of the alphabet used, so that the benefits outweigh the complications by a larger amount as the range of $l$ values increases.

It is necessary to keep the possible key values uniformly distributed, so that Eve can't obtain any advantage from knowledge of the nonuniform distribution. To
do this it will be necessary to make sure that the spectrum of signal and idler values is broader than the alphabet of values actually used for the key, since
the nonorthogonality of the superposition states means that measurements can broaden the range of outgoing values (see section \ref{evesection}).

It should be noted that photons are lost from the setup only once, in the filtering after the crystal; the spiral before the crystal simply redistributes the energy among the available modes through interference effects, with no net loss of photons or energy. As a result, event rates should be at levels comparable to other entangled photon protocols.

For illustration purposes, we will often restrict ourselves to the case $N=8$. Assume that the pump spectrum is broad enough to be approximately flat over a
sufficient span to produce signal and idler OAM values of uniform probability over the range $F_{m_0}$ to $F_{m_0+N-1}$, for some $m_0$. The outcomes for
$\mathbb{L}$-type detection (OAM eigenstates) that will be used for key generation by Alice and Bob are then simply $|F_{m_0}\rangle ,|F_{m_0+1}\rangle, \dots
|F_{m_0+N-1}\rangle$. For example, if the values $N=8$, $m_0=2$ are chosen, then the utilized states are
\begin{eqnarray}|l_A\rangle ,\; |l_B\rangle &=& \left\{ |F_2\rangle , |F_3\rangle , |F_4\rangle , \dots ,
|F_9\rangle \right\} \\ &=& \left\{  |2\rangle ,  |3\rangle ,|5\rangle , |8\rangle , |13\rangle ,
|21\rangle ,|34\rangle , |55\rangle \right\} .\nonumber\end{eqnarray}

The outcomes for $\mathbb{D}$-type detection (two-fold OAM superposition states) used by Alice and Bob for key generation run from \begin{equation}
|S_{m_0}\rangle = {1\over \sqrt{2}}\left\{ |F_{m_0-1}\rangle +|F_{m_0+1} \rangle\right\} \end{equation} to
 \begin{equation} |S_{m_0+N-1}\rangle = {1\over \sqrt{2}}\left\{ |F_{m_0+N-2}\rangle +|F_{m_0+N} \rangle\right\}  .\end{equation}

Detection either of the states $|F_n\rangle $ or $|S_n\rangle$ by Alice will correspond to a key value $K=F_n$. Unfortunately, Alice and Bob will not necessarily
agree on the same key unless further classical information is exchanged to reconcile their values. One possible method for reconciliation is discussed in
appendix A.

\section{Effect of eavesdropping: qualitative discussion} \label{evesection}

\begin{figure}
\begin{center}
\subfigure[]{
\includegraphics[height=1.0in]{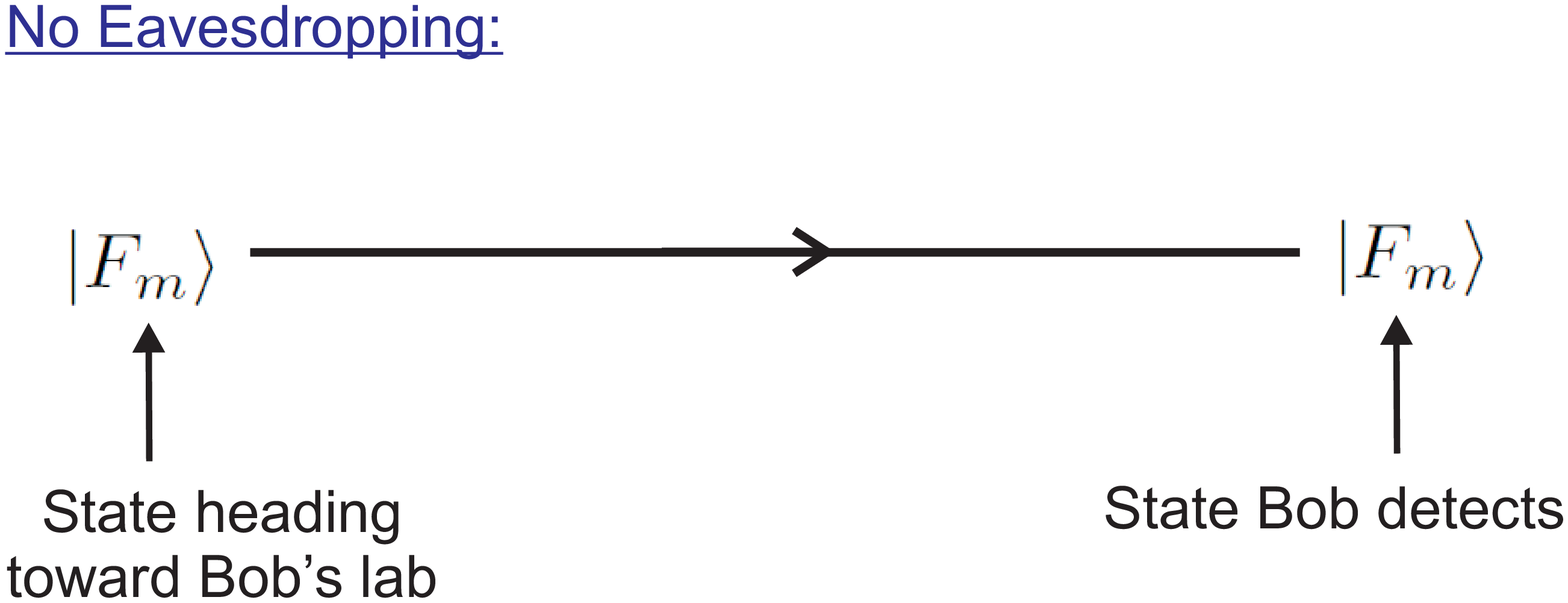}}
\subfigure[]{
\includegraphics[height=2.0in]{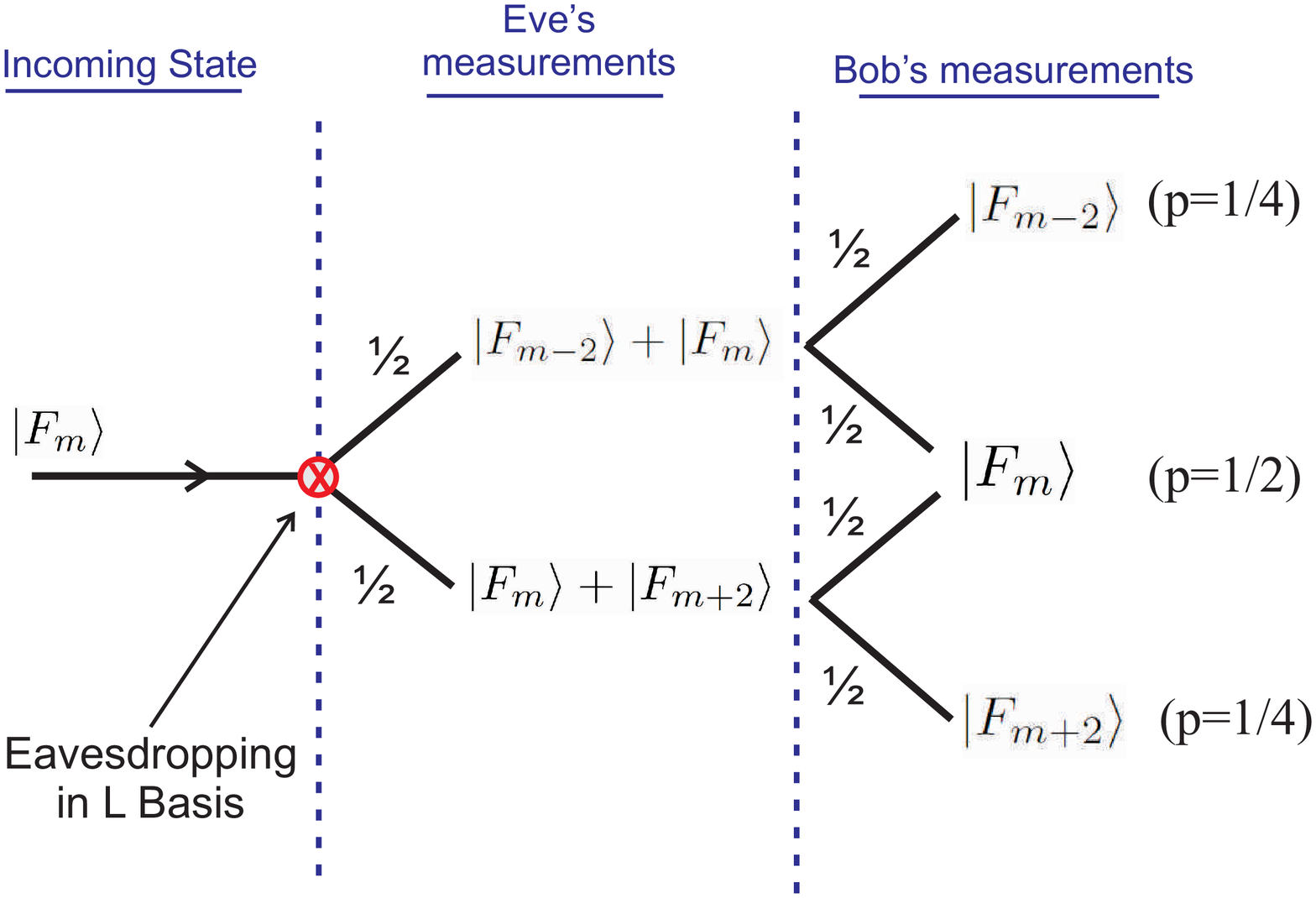}}
\caption{Outcome probabilities for eigenstates. (a) When there is no eavesdropping and Bob measures in the $\mathbb{L}$ basis, an incoming eigenstate should be detected correctly $100\%$ of the time. (b) When Eve measures in the $\mathbb{D}$ basis, each eigenstates can result in
two different superposition detections. If she sends one of these superpositions on to Bob, the net result is that there are now three eigenstates that he could detect.}\label{eigoutcomesfig}
\label{outcome_eig_fig}\end{center}
\end{figure}

If an eavesdropper is acting on the photon heading to Bob, she does not know which type of detection ($\mathbb{D}$ or $\mathbb{L}$) will occur in Alice's and
Bob's labs. If Alice detects an eigenstate, then the state arriving at Bob's end should be a superposition, whereas if Alice detects a superposition then the
state heading toward Bob should be an eigenstate. If Eve makes a $\mathbb{D}$-type measurement when Bob's photon is in an $\mathbb{L}$ state or if she makes an
$\mathbb{L}$-type measurement when Bob's photon is in a $\mathbb{D}$ state, a disturbance is made in the statistical distribution of Alice and Bob's joint
outcomes, which will become apparent when he compares a random subset of his trials with Alice's. In more detail:

(i) Suppose Eve makes a $\mathbb{D}$-type measurement on a photon which is actually in the eigenstate $|F_m\rangle$. She will detect one of the two
superpositions $|F_m\rangle+|F_{m-2}\rangle$ or $|F_m\rangle+|F_{m+2}\rangle$, each with $50\%$ probability, and send on a copy of it. If Bob receives one of
these superpositions and makes an $\mathbb{L}$ measurement, he will see one of the values $F_m$, $F_{m-2}$, or $F_{m+2}$, with respective probabilities of
$1\over 2$, $1\over 4$, $1\over 4$. In Eve's absence, he should only see $F_m$ with $100\%$ probability (see Fig. \ref{outcome_eig_fig}).

\begin{figure}[ht!]
\begin{center}
\subfigure[]{
\includegraphics[height=1.0in]{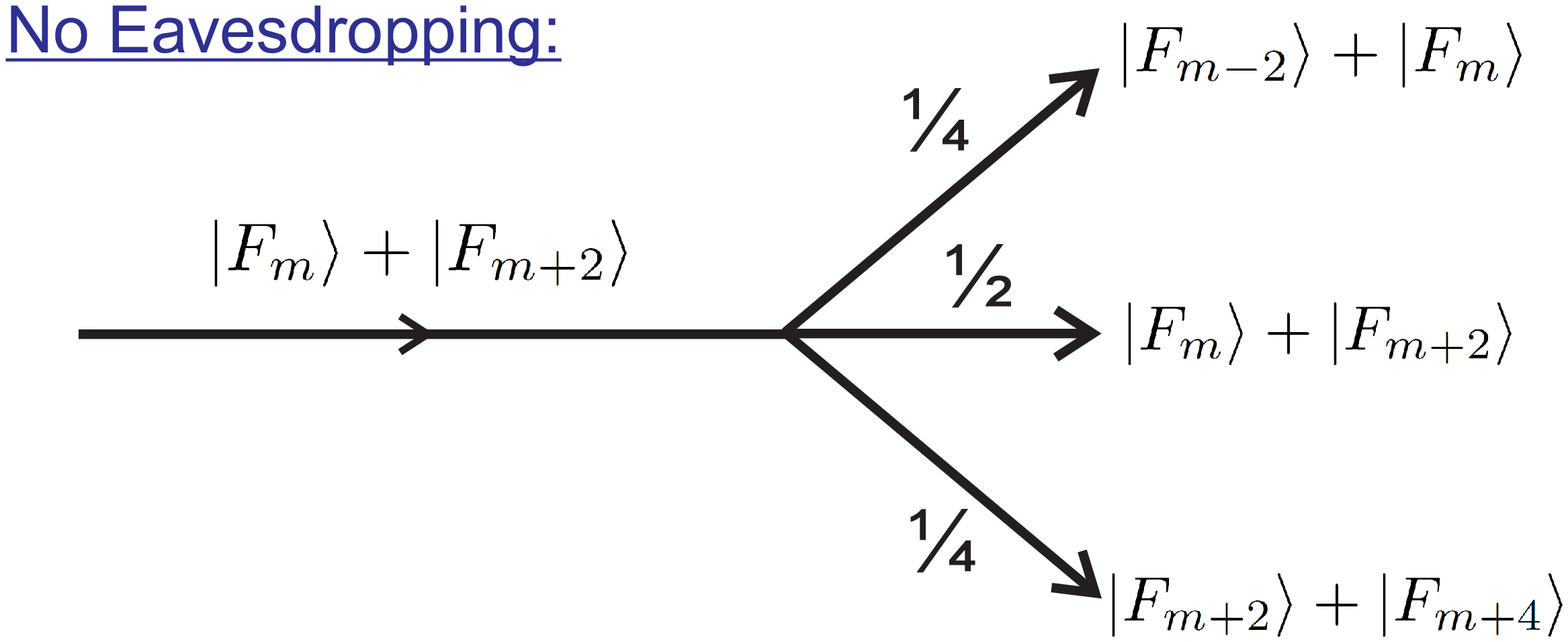}}
\subfigure[]{
\includegraphics[height=2.0in]{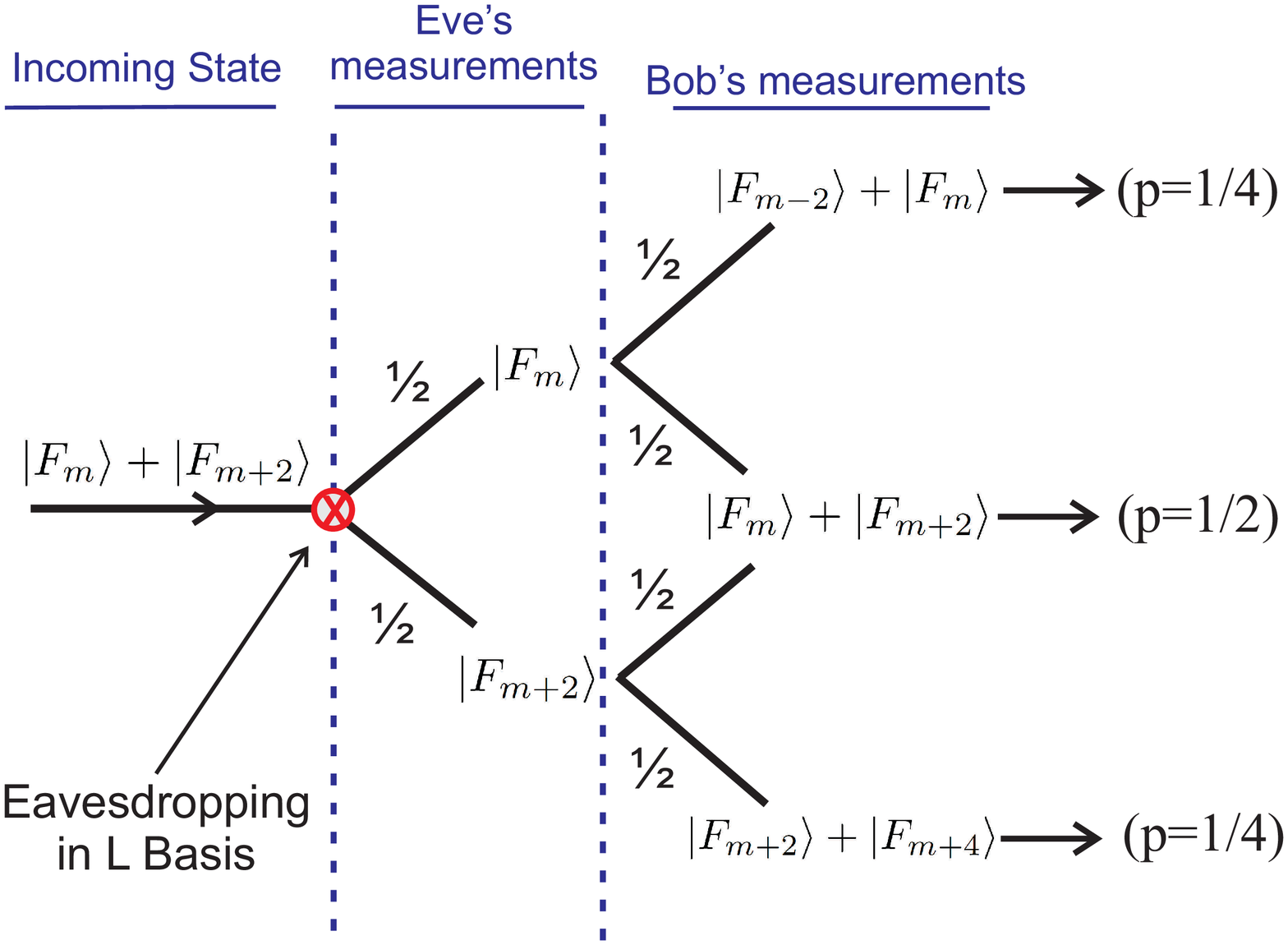}}
\caption{Outcome probabilities for superposition states. (a) When there is no eavesdropping, an incoming superposition state leads to three
possible outcomes when Bob makes a measurement in the superposition ($\mathbb{D}$) basis: the state can be identified correctly with probability
${1\over 2}$, or it can be misidentified as one of the other two allowed superpositions that have nonzero overlap with it. (b) When Eve measures in the
$\mathbb{L}$ basis, each of the two possible eigenstates that result
can lead to two different superpositions. The net result is that there are three outcomes that have nonzero overlap with the two eigenstates,
so that the same probabilities occur as in figure (a), unless an addition is made to the apparatus (section \ref{supersection}).}\label{outcomesfig}
\end{center}
\end{figure}

(ii) On the other hand, suppose Eve makes an $\mathbb{L}$-type measurement on a photon which is actually in the superposition state $|F_m\rangle +|F_{m-2}\rangle$. She
will detect one of the two eigenstates $|F_m\rangle$ or $|F_{m-2}\rangle$, each with $50\%$ probability, and send on a copy of it. If Bob receives one of these
eigenstates and makes a $\mathbb{D}$ measurement, he will see one of the superpositions $|F_m\rangle+|F_{m-2}\rangle$, $|F_m\rangle+|F_{m+2}\rangle$, or
$|F_{m-2}\rangle+|F_{m-4}\rangle$, with respective probabilities of $1\over 2$, $1\over 4$, $1\over 4$ (see fig. \ref{outcomesfig}).  Thus, it seems that Bob sees the same result whether Eve interferes or not. However, by adding an additional interferometric element to the setup, we will see (section \ref{supersection}) that Bob will be able to distinguish between the two cases.

It will be desirable to arrange the probabilities of the possible outcomes into a matrix. However, as seen above, the process of measurement can cause the range of states
present in the system to spread. States that are within the range of our alphabet can lead to measurements outside the allowed range (for example,
the in-range state $|S_{m_0+N-1} \rangle$ can be measured as the out-of-range $|S_{m_0+N} \rangle$, since these two states have nonzero overlap). Similarly, states that are initially outside the desired
range can lead to in-range measurements. This spreading is greater when Eve is present and introducing additional measurements. So to account for this and to
maintain the uniform distribution of key values, we allow states beyond the desired range to propagate in the system and include additional rows and columns in
the probability matrix to describe the probability that the measured state is out of range of the allowed alphabet. This will be seen explicitly in section
\ref{probsection} and Appendix B. Note that we only need to include events in which at least one of the legitimate participants sees a value out of range; we can
exclude events where the measurements are out of range for both of them.

\section{Discriminating superposition states}\label{supersection}

\begin{figure}
\begin{center}
\includegraphics[scale=.3]{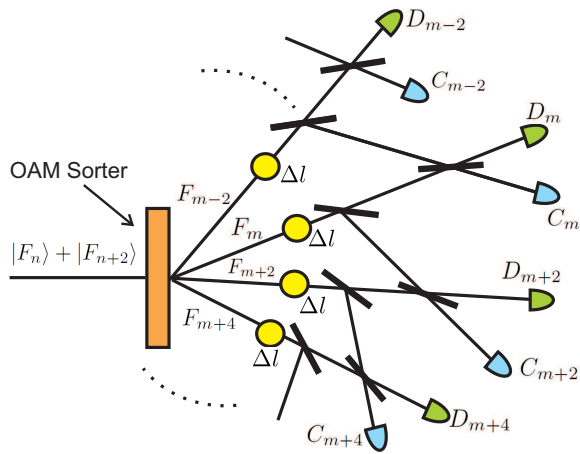}
\caption{A portion of a detection unit for statistically detecting superpositions of OAM states, analogous to the polarization version of
the previous figure. The sorter separates different OAM values,
which are then shifted to zero OAM by spiral wave plates or holograms (the yellow circles).  From the top downward, the OAM shifters shown
change the incoming OAM value by $\Delta l=-F_{m-2}, -F_{m}, -F_{m+2},  -F_{m+4}$. Superpositions of the form $|F_m\rangle +|F_{m+2}\rangle$ cause constructive interference at the $C_m$
detectors and destructive interference at the $D_m$ detectors, while OAM eigenstates lead to equal detection rates at both types.} \label{CDfig}
\end{center}
\end{figure}

The discrimination of OAM eigenstates in the $\mathbb{L}$ basis is straightforward and simply requires an OAM sorter. Discrimination of superposition states in the $\mathbb{D}$ basis is more complicated.
As shown in \cite{simon2}, the setup of figure \ref{CDfig} is capable of sorting the superposition states. A pair of photon-counting detectors $C_n$ and $D_n$ is
used at the output ports of the final nonpolarizing beam splitters. If $C_n$ fires during the key-generating trials, we count that as an $|S_{n-1}\rangle $
detection. Due to destructive interference, $D_n$ should not fire for $|S_{n-1}\rangle $ input, so its firing will count as an $|F_n\rangle $ detection. Then the scheme
of Appendix A is used to reconcile Alice's and Bob's trials by classical information exchange in order to arrive at an unambiguously agreed-upon key. During the
security checks, we then look at the distribution of counts in $C_n$ and $D_n$ separately in order to detect eavesdropper-induced deviations from the expected
probability distributions. In order to achieve the indistinguishability required for interference, the OAM of each photon is shifted to zero after the sorting
by a spiral phase plate or by other means. Measurements with detectors $C_n$ and $D_n$, respectively, are equivalent to looking for nonzero
projections onto the states
\begin{eqnarray}|C_n\rangle &=& {i\over \sqrt{2}} \left( |F_n\rangle +|F_{n-2}\rangle \right) \; =\; i|S_{n-1}\rangle\\
|D_n\rangle &=& {1\over \sqrt{2}} \left( |F_n\rangle -|F_{n-2}\rangle \right)  .
\end{eqnarray} These two sets of states are also mutually nonorthogonal: $\langle C_n|D_m\rangle =-{i\over 2}\left( \delta_{m,n-2}-\delta_{m,n+2}\right) .$ )

There are then multiple possible output states for Alice and Bob. Each can measure in the $\mathbb{L}$ basis and obtain one of the $|F_n\rangle$ states, or they
can measure in the $\mathbb{D}$ basis and register one of the $|C_n\rangle$ or $|D_n\rangle $ states. Their joint output probabilities for these states can then
be found by taking the incoming state, applying the appropriate projection operator ($|F_n\rangle\langle F_n|$, $|C_n\rangle\langle C_n|$, $|D_n\rangle\langle
D_n|$), and then taking the inner product of the resulting projection with the initial state. The effect of Eve's intervention on Bob's channel can be dealt with
by inserting similar additional projection operators representing Eve's measurements. These extra projection operators tend to spread the probability
distributions out, causing them to be nonzero for combinations of outcomes that previously had vanishing probabilities. As a result, Eve's actions can be
revealed by these alterations in the outcome probabilities.

\section{Mutual Information and Security}\label{probsection}

For the detection events described in the previous sections, we may now construct the probability distributions.

We begin by making some definitions. $P_{ij}$ will be the joint probability that Alice measures outcome $i$ and Bob measures outcome $j$ in the absence of
eavesdropping. (In other words, these are the \emph{expected} probabilities.)  $P_{Eij}$ will be the \emph{observed} probabilities for the same outcomes in the
presence of eavesdropping. The index $i$ labels Alice's outcomes. $i=1$ represents an $\mathbb{L}$-type measurement with with the measured value out of range of
the desired alphabet, while values $2\le i\le N+1$ correspond to the event that Alice measured in the $\mathbb{L}$ basis and found state $|F_{m_0+i-2}\rangle$;
similarly, $1\leq j \leq N+1$ represents Bob detecting an out-of-range value or measuring state $|F_{m_0+j-2}\rangle$ in the eigenbasis. Value $i=N+2$ represents
Alice making an out-of-range detection in the $\mathbb{D}$ basis, while values of $i$ in the range $N+3\leq i\leq 3N+2$ represent Alice measuring in the diagonal
superposition basis and seeing an event in the $C$ and $D$ detectors. Specifically, $i=N+2k+3$ represents the firing of $C_{m_0+k}$, and $i=N+2k+4$ represents
the firing of $D_{m_0+k}$, for $k=0, \dots N-1$; similarly for Bob with $j$ values in the same range. $P_i^{(A)}$ and $P_j^{(B)}$ will be the expected marginal
probabilities, $P_i^{(A)}=\sum_jP_{ij}$ and $P_j^{(B)}=\sum_iP_{ij}$, with similar definitions for $P_{Ei}^{(A)}$ and $P_{Ej}^{(B)}$.

First, we give the expected probabilities for the states in the two beams (before Bob's measurement) in the absence of eavesdropping. Columns label Alice's
measurements, while rows label Bob's. The matrix of outcome probabilities then has the form \begin{equation}P_0={1\over 4}\left( \begin{array}{cc} L_0 & C_0 \\
C_0^T & D_0
\end{array} \right) \label{probmatrix},\end{equation} where explicit expressions for the submatrices $L_0$, $C_0$, and $D_0$ may be found in Appendix B.

The matrices $L_0$ and $D_0$ represent, respectively, the events on which both Alice and Bob measured in the $\mathbb{L}$ basis or both in the $\mathbb{D}$ basis. $C_0$
represents events in which Alice and Bob measured in different bases, one $\mathbb{L}$ and one $\mathbb{D}$. The factor of ${1\over 4}$ in eq. \ref{probmatrix} is due to the
fact that each of the four combinations of detection type ($\mathbb{LL}$, $\mathbb{DD}$, $\mathbb{LD}$, and $\mathbb{DL}$) has a probability of $1\over 4$ in a
given trial.

\begin{figure}
\begin{center}
\includegraphics[scale=.33]{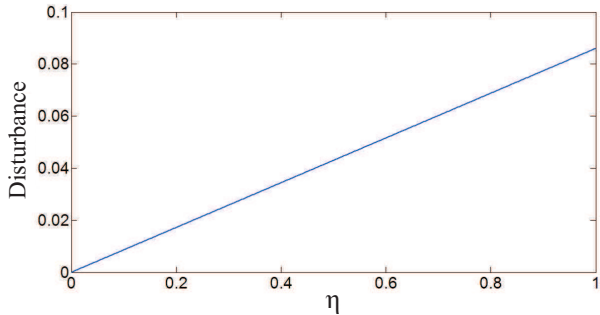}
\caption{The disturbance to the probability distribution, plotted against the eavesdropping fraction, $\eta$.}
\label{disturbplotfig}
\end{center}
\end{figure}

Now let $\eta$ be the proportion of trials on which Eve eavesdrops, i.e. the fraction of the photons on which she makes measurements. Assume she also randomly
measures (with equal probabilities) in the $\mathbb{L}$ or $\mathbb{D}$ basis. Then the probability matrix for Alice and Bob's outcomes will change according to:
\begin{equation} P_0\to P=(1-\eta )P_0 +{\eta} P_E ,\end{equation} where
\begin{equation}P_E ={1\over 4}\left( \begin{array}{cc} L^\prime & C^\prime \\ F^\prime & D^\prime \end{array} .\right)\end{equation} The entries in $P_E$ give the probability that Eve's actions will induce a change in the measured value, given that she intervened on that particular trial.
The new submatrices $L^\prime$, $C^\prime$, $D^\prime$, and $F^\prime$  may be found in Appendix B. We assume for simplicity that Eve only acts on Bob's channel,
not Alice's. The more general case can be treated in a similar manner. Note that $F^\prime$ no longer needs to be equal to $C^{\prime T}$ due to the asymmetry
introduced by this assumption. We define the change in the probability matrix due to eavesdropping,
\begin{eqnarray}\Delta P&=& P-P_0\\
& =& {\eta \over 4}\left( \begin{array}{cc} L^\prime -L_0 & C^\prime -C_0 \\ F^\prime -C_0^T& D^\prime -D_0 \end{array} .\right) .\label{delP}\end{eqnarray} The {\it disturbance} \cite{lutken96} introduced by the eavesdropper can be used as a measure of how much effect she is having on the outcomes of the communication. The disturbance ${\cal D}$ is defined to be
\begin{equation}{\cal D} =\sqrt{ \sum_{ij} \left(\Delta P_{ij}\right)^2}.\label{disturbdef}\end{equation}
For binary protocols like BB84 the disturbance simply equals the eavesdropper-induced error rate, so ${\cal D}$ is an appropriate generalization to protocols for
which there are more than two possible outcomes per measurement. Using the probability matrices of Appendix B, ${\cal D}$ can be readily calculated. From eqs.
\ref{delP} and \ref{disturbdef} it should clearly be linear in the eavesdropping fraction $\eta$, as verified in the plot of fig. \ref{disturbplotfig}. The
disturbance reaches similar values to those found in \cite{lutken96} for the BB84 protocol.

From the probability distributions, the mutual information shared by Alice and Bob can also be computed:
\begin{eqnarray}I_{AB}&=&H_A+H_B-H_{AB}\label{I_mut1} \\ &= & -\sum_i P_{A,i}\log_2P_{A,i} -\sum_j P_{B,j}\log_2P_{B,j}\nonumber \\
& & \qquad\quad +\sum_{ij} P_{i,j}\log_2P_{i,j}.\label{I_mut2}\end{eqnarray}
The information per trial gained by Eve can also be found.
First note that the probability that she measures in the correct basis is $1\over 2$. If she guesses the correct basis then she can determine the correct value for Bob with certainty; if she guesses the wrong basis, she only has a $50\%$ chance of determining the correct value. So on a given trial, her probability of obtaining the correct value is $3\over 4$. Therefore, her information gain per trial (her average mutual information with Alice) is:
 \begin{eqnarray}I_{AE} &=&\left( \mbox{prob. that Eve listens in on trial}\right)
\\ & & \qquad \cdot \left( \mbox{probability trial is not discarded} \right)\nonumber \\
& & \qquad  \cdot \left( \mbox{info. gained per eavesdropping}\right)\nonumber \\
&=& \eta \;\cdot \; r(\eta )\; \cdot \; I_{AB}
,\end{eqnarray} where the fraction $r(\eta )$ of trials retained is given in Appendix B.

\begin{figure}
\begin{center}
\includegraphics[scale=.3]{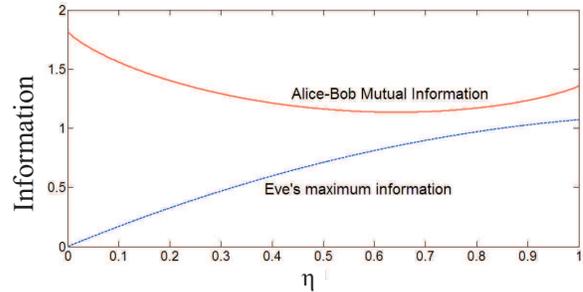}
\caption{The mutual information (upper red curve) shared between Alice and Bob, and the information gain per trial by Eve (lower blue curve) as functions
of $\eta$. 
}
\label{info1fig}
\end{center}
\end{figure}

\begin{figure}
\begin{center}
\includegraphics[scale=.32]{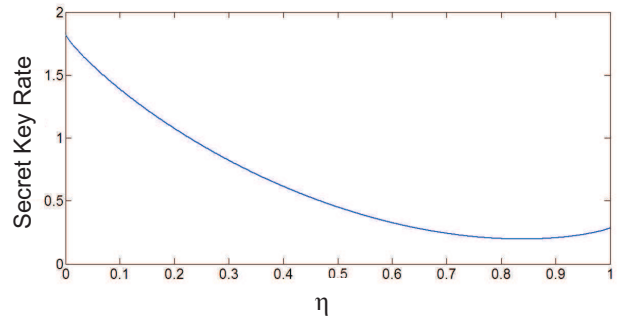}
\caption{The secret key generation rate $\kappa$ versus eavesdropping fraction $\eta$. }
\label{keyfig}
\end{center}
\end{figure}

\begin{figure}
\begin{center}
\includegraphics[scale=.32]{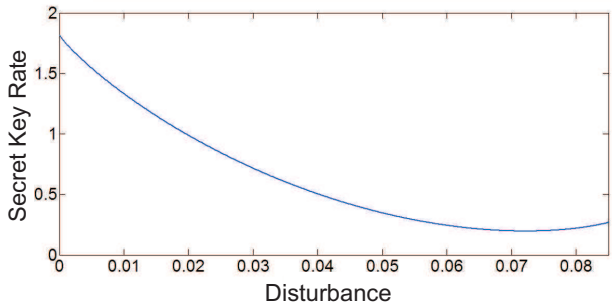}
\caption{The secret key rate $\kappa$ as function of disturbance, ${\cal D}$.  }
\label{infordistfig}
\end{center}
\end{figure}

These two information measures are plotted in fig. \ref{info1fig} as a function of eavesdropping fraction $\eta$. It can be seen that the mutual information per
trial gained by Eve is always less than the mutual information shared between Alice and Bob, so that a secure key can always be distilled from the exchange. The
secret key rate $K= I_{AB}-I_{AE}$ is shown in fig. \ref{keyfig} as a function of $\eta$ and as a function of disturbance in fig. \ref{infordistfig}. The slight upturn
in the Alice-Bob mutual information at large $\eta$ in Fig. \ref{info1fig} seems to be due to the fact that Eve's interference causes the outcomes to be more
spread out (there are more nonzero entries in $P_E$ than in $P_0$), and more uniform distributions lead to increased information gains per measurement. Due to
the concave shape of the binary entropy function, such an effect can also happen in schemes using binary qubits, if the error rate is able to exceed $50\%$.

\section{Conclusions}\label{conclusionsection}

In this paper, we have constructed the joint probability distributions for Alice and Bob's measurement outcomes in the Fibonacci protocol, both in the
absence of eavesdropping and for the case of simple intercept-resend attacks. For the case of three bits per photon, it has been demonstrated that a secure key can be
distilled and that the eavesdropper's presence can be revealed. For larger alphabets (more key bits per photon), the mutual information shared by Alice and Bob increases logarithmically with alphabet size, while the amount of classical information that needs to be exchanged stays constant. It remains for future investigation to see
how the protocol behaves under more sophisticated eavesdropping attacks.

\section*{Acknowledgements} This research was supported by the DARPA QUINESS program through US Army Research Office award
W31P4Q-12-1-0015

\section*{Appendix A: Reconciliation}

In order for Alice and Bob to determine each other's values they must exchange additional information on a classical side channel.  Here, a brief description is
given of one way to do this. Other methods are also possible. It is necessary to make sure that no eavesdropper can obtain the key from the classical exchange
alone.

If Eve intercepts both the classical and quantum exchanges, and if she can store the photon from the quantum channel until she has read the classical channel,
then she has complete information about the bit, just as Bob does. However, the disturbances she introduces will then signal her presence, causing the tainted
key to be discarded. This is identical to the case in the BB84 or Ekert protocols.

However, in the Fibonacci case a \emph{larger} classical exchange is required, which reduces the average amount of information about the key that remains secret.
This reduces the key generation capacity per photon, partially canceling the advantages from the larger Hilbert space. But even in the worst case the classical
exchange can be chosen such that the amount of classically revealed information remains less than the amount of information generated by the quantum exchange,
allowing the distillation of a secure key \cite{cziszar}. Furthermore, the amount of revealed information is independent of $N$, while the total information from
the quantum exchange increases with increasing alphabet size like $\log_2N$; thus this information leakage becomes more and more negligible with increasing $N$,
as compared to the total information.

When both experimenters measure in the $\mathbb{L}$ (eigenstate) basis, they should (in the absence of eavesdropping) always receive adjacent Fibonacci
numbers, say $F_n$ and $F_{n+1}$, although initially neither of them knows whether they have the lower or higher value in the pair. In order for them to
determine this, they must each exchange one bit of information over the classical channel.

One possible way to do this (illustrated for the case $N=8$, $m_0=2$), is for Alice first to send either a $0$ or $1$ to Bob in the following manner:
\begin{equation}\begin{array}{r|c|c|c|c|c|c|c|c|c} \mbox{Alice has:}
& 2 & 3 & 5 & 8 & 13 & 21 & 34& 55\\ \hline \mbox{Alice sends:} & 0 & 1 & 1 & 0 & 0 & 1 & 1  &0  \end{array}\end{equation} Once Bob receives this information he
can determine Alice's value, since he already knows it has to be one of the two values adjacent to his. Alice's value can then be used as the key value. For
arbitrary $N$, the probability of guessing the correct value based on the classical exchange alone decreases with increasing $N$: $P={2\over N}$.

Half of the time, Alice and Bob make opposite types of measurements, one $\mathbb{L}$ and one $\mathbb{D}$. Suppose, for example, Alice measures the value $F_5=8$ in
the $\mathbb{L}$ basis, while Bob measures in $\mathbb{D}$. In principle, he should receive the superposition state ${1\over \sqrt{2}} \left( |5\rangle +|13
\rangle\right) $; however, since non-orthogonal states can not be uniquely distinguished, there is a $25\%$ chance that Bob will instead measure the
superposition  ${1\over \sqrt{2}} \left( |2\rangle +|5 \rangle\right) $ and $25\%$ that he will find ${1\over \sqrt{2}}\left( |13\rangle +|34 \rangle\right) $.
To arrive at an unambiguous value shared by both participants, there are several possible procedures. One possibility (again described for the case $N=8$,
$m_0=2$) is for Alice to send Bob {\it two} bits of information, according to the following table:
\begin{equation}\begin{array}{r|c|c|c|c|c|c|c|c|c} \mbox{Alice has:} & 2 & 3 & 5 & 8 & 13 & 21 & 34& 55\\ \hline
\mbox{Alice sends:} & 01 & 10 & 00 & 01 & 10 & 00 & 01  &10  \end{array}\end{equation} Bob then knows Alice's value unambiguously, so that they can agree to use
her value as the key segment; again, there is no need for Bob to send any classical information in this case. As $N$ increases, the number of key bits generated
per photon grows but the amount of classical information exchanged remains fixed.

Finally, Alice and Bob can both make a $\mathbb{D}$ measurement. Suppose that Alice sends two bits of classical information, which
could be: \begin{equation}\begin{array}{r|c|c|c|c|c|c|c|c|c|c|c} \mbox{Alice has:} & S_1 & S_2 & S_3 & S_4 & S_5 & S_6 & S_8& S_9& S_{10} & S_{11} & \dots\\
\hline \mbox{Alice sends:} & 00 & 00 & 01 & 01 & 10 & 10 & 11  &11  & 00 & 00 & \dots \end{array}\end{equation} Examination of the probability matrices (eqs.
\ref{L0eq}-\ref{D0eq}) makes clear that, armed with knowledge of his own superposition, Bob can unambiguously determine Alice's value, while an outsider
eavesdropper on the classical channel again has a decreasing probability of guessing the correct value as N increases. Once Alice and Bob agree that Alice has
superposition $|S_n\rangle $, they then adopt the value $F_n$ as the key value.

\section*{Appendix B: Probability Matrices}

\begin{figure}
\begin{center}
\includegraphics[scale=.32]{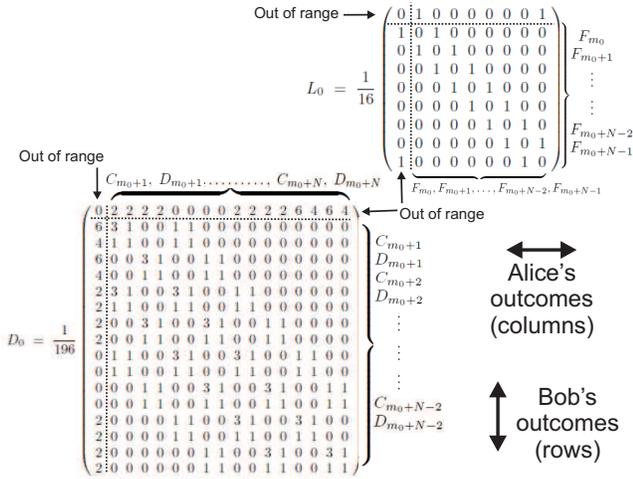}
\caption{The labeling of rows and columns is shown for the probability submatrices. Different rows correspond to different outcomes for Bob,
while columns correspond to Alice's outcomes. The first row and column label detection of outcomes not included in the range
of the alphabet used for key generation. The remaining rows and columns of $L_0$ label allowed eigenstates, and those of $D_0$
alternately label $C_n$ and $D_n$ detection events signalling allowed superposition states. Similarly, $C_0$  has eigenstate events running
horizontally, with $C_n$ and $D_n$ events vertically. $L_0$, $C_0$, and $D_0$ are respectively $(N+1)\times (N+1)$, $(N+1)\times (2N+1)$, and
$(2N+1)\times (2N+1)$ matrices.} \label{matrixfig}
\end{center}
\end{figure}

Here we give explicit expressions for the joint probability distributions seen by Alice and Bob for their outcomes, with and without eavesdropping. See section
\ref{probsection} for the relevant definitions. The labeling of the rows and columns is shown in fig. \ref{matrixfig}.

To compute the probabilities in the absence of eavesdropping, first consider the projections of the biphoton state $|\psi\rangle$ after the crystal (eq.
\ref{initialstate}) onto the states that can be detected at the various detectors on Alice's side:
\begin{eqnarray}|\Psi_{F_n}^A\rangle &=& |F_n\rangle_A\; _A\langle F_n|\psi\rangle \\ & =& {1\over \sqrt{2}} |F_n\rangle_A \; \big(
|F_{n-1}\rangle_B+ |F_{n+1}\rangle_B\big) \\
|\Psi_{D_n}^A \rangle &=& |D_n\rangle_A\; _A\langle D_n|\psi\rangle \\ & =& -{i\over 2} |D_n\rangle_A \\ & & \quad \times \big(
|F_{n-3}\rangle_B++2|F_{n-1}\rangle_B+ |F_{n+1}\rangle_B\big)\nonumber \\
|\Psi_{C_n}^A \rangle &=& |C_n\rangle_A\; _A\langle C_n|\psi\rangle \\ & =& {1\over 2} |C_n\rangle_A \; \big( |F_{n+1}\rangle_B-
|F_{n-3}\rangle_B\big) ,
\end{eqnarray}
with similar expressions for the states $|\Psi_{F_n}^B\rangle$, $|\Psi_{C_n}^B\rangle$, $|\Psi_{D_n}^B\rangle$ on Bob's side. Then (up to an overall constant
that can be fixed by the requiring the probabilities to sum to one) the matrix for the probabilities of joint detections by Alice and Bob can be built up entry
by entry by computing the various products $\langle \Psi_{X_n}^A| \Psi_{Y_m}^B \rangle $, for $X,Y=C,D,F$. We give here the results for $N=8$; the generalization
to larger $N$ is straightforward. The results are:

\scriptsize
\begin{widetext}
\begin{eqnarray} L_0& =& {1\over{16}}\left( \begin{array}{ccccccccc}
0 & 1& 0& 0& 0& 0& 0& 0 &  1\\
1 & 0 & 1 & 0& 0& 0& 0& 0& 0\\
0 & 1 & 0 & 1& 0& 0& 0& 0& 0\\
0 & 0 & 1 & 0 & 1& 0& 0& 0& 0\\
0 & 0& 0 & 1 & 0 & 1& 0& 0& 0 \\
0 & 0& 0& 0 & 1 & 0 & 1& 0& 0\\
0 & 0& 0& 0& 0 & 1 & 0 & 1& 0\\
0 & 0& 0& 0& 0& 0 & 1 & 0 &1\\
1 & 0& 0& 0& 0& 0& 0 & 1 &0\\
\end{array} \right) \qquad
C_0 \; =\; {1\over {72}}\left( \begin{array}{ccccccccccccccccc}
0 &1 & 1& 1& 1& 0& 0& 0 & 0 &0 & 0& 0& 0& 1& 1& 1 & 1\\
2 &4 & 0& 0& 0& 1& 1& 0 & 0 &0 & 0& 0& 0& 0& 0& 0 & 0\\
2 &0 & 0& 4& 0& 0& 0& 1 & 1 &0 & 0& 0& 0& 0& 0& 0 & 0\\
0 &1 & 1& 0& 0& 4& 0& 0 & 0 &1 & 1& 0& 0& 0& 0& 0 & 0\\
0 &0 & 0& 1& 1& 0& 0& 4 & 0 &0 & 0& 1& 1& 0& 0& 0 & 0\\
0 &0 & 0& 0& 0& 1& 1& 0 & 0 &4 & 0& 0& 0& 1& 1& 0 & 0\\
0 &0 & 0& 0& 0& 0& 0& 1 & 1 &0 & 0& 4& 0& 0& 0& 1 & 1\\
2 &0 & 0& 0& 0& 0& 0& 0 & 0 &1 & 1& 0& 0& 4& 0& 0 & 0\\
2 &0 & 0& 0& 0& 0& 0& 0 & 0 &0 & 0& 1& 1& 0& 0& 4 & 0\\
\end{array} \right) \label{L0eq} \\
D_0 &=& {1\over {196}}\left( \begin{array}{ccccccccccccccccc}
0 &2 & 2& 2& 2& 0& 0& 0 & 0 &2 & 2& 2& 2& 6& 4& 6 & 4\\
6 &3 & 1& 0& 0& 1& 1& 0 & 0 &0 & 0& 0& 0& 0& 0& 0 & 0\\
4 &1 & 1& 0& 0& 1& 1& 0 & 0 &0 & 0& 0& 0& 0& 0& 0 & 0\\
6 &0 & 0& 3& 1& 0& 0& 1 & 1 &0 & 0& 0& 0& 0& 0& 0 & 0\\
4 &0 & 0& 1& 1& 0& 0& 1 & 1 &0 & 0& 0& 0& 0& 0& 0 & 0\\
2 &3 & 1& 0& 0& 3& 1& 0 & 0 &1 & 1& 0& 0& 0& 0& 0 & 0\\
2 &1 & 1& 0& 0& 1& 1& 0 & 0 &1 & 1& 0& 0& 0& 0& 0 & 0\\
2 &0 & 0& 3& 1& 0& 0& 3 & 1 &0 & 0& 1& 1& 0& 0& 0 & 0\\
2 &0 & 0& 1& 1& 0& 0& 1 & 1 &0 & 0& 1& 1& 0& 0& 0 & 0\\
0 &1 & 1& 0& 0& 3& 1& 0 & 0 &3 & 1& 0& 0& 1& 1& 0 & 0\\
0 &1 & 1& 0& 0& 1& 1& 0 & 0 &1 & 1& 0& 0& 1& 1& 0 & 0\\
0 &0 & 0& 1& 1& 0& 0& 3 & 1 &0 & 0& 3& 1& 0& 0& 1 & 1\\
0 &0 & 0& 1& 1& 0& 0& 1 & 1 &0 & 0& 1& 1& 0& 0& 1 & 1\\
2 &0 & 0& 0& 0& 1& 1& 0 & 0 &3 & 1& 0& 0& 3& 1& 0 & 0\\
2 &0 & 0& 0& 0& 1& 1& 0 & 0 &1 & 1& 0& 0& 1& 1& 0 & 0\\
2 &0 & 0& 0& 0& 0& 0& 1 & 1 &0 & 0& 3& 1& 0& 0& 3 & 1\\
2 &0 & 0& 0& 0& 0& 0& 1 & 1 &0 & 0& 1& 1& 0& 0& 1 & 1\\
\end{array} \right)\label{D0eq}
\end{eqnarray}
\end{widetext}

%

\normalsize Eve's measurements will alter Bob's outcome probabilities. If $\eta$ be the proportion of trials she intercepts,  then the probability matrix for
Alice and Bob's outcomes will change according to $P_0\to P=\eta P_E+(1-\eta )P_0 ,$ where $P_E$ describes the change in probability due to Eve's intervention on
a given trial. The various joint probabilities are computed as before, but with additional projection operators inserted into Bob's line to represent Eve's
action. For example, when Alice and Bob measure in the $\mathbb{L}$ basis and Eve measures in the $\mathbb{D}$ basis, the probability that Alice detects $F_p$
and Bob detects $F_n$ has extra terms added to it of the form \begin{eqnarray}& & \sum_m\left\{ \langle\Psi_{F_p}^A| \Psi_{C_m}^B \rangle |\langle C_m|F_n\rangle
|^2 \right. \\ & & \left.\qquad+  \langle\Psi_{F_p}^A| \Psi_{D_m}^B \rangle |\langle D_m|F_n\rangle |^2 \right\} ;\nonumber
\end{eqnarray} these extra terms correspond to the possible outcomes of Eve's measurement. Similar expressions apply
to find the rest of the probabilities. The net result (for $N=8$) is: $P_E ={1\over 4}\left( \begin{array}{cc} L^\prime & C^\prime \\ F^\prime & D^\prime
\end{array} \right) ,$ where the new submatrices are \scriptsize
\begin{widetext}
\begin{equation} L^\prime  = {1\over{122}}\left( \begin{array}{ccccccccc}
0 & 1& 0& 0& 1& 1& 6& 6 &  11\\
11 & 0 & 1 & 0& 0& 0& 0& 0& 0\\
6 & 5 & 0 & 1& 0& 0& 0& 0& 0\\
6 & 0 & 5 & 0 & 1& 0& 0& 0& 0\\
1 & 5& 0 & 5 & 0 & 1& 0& 0& 0 \\
1 & 0& 5& 0 & 5 & 0 & 1& 0& 0\\
0 & 1& 0& 5& 0 & 5 & 0 & 1& 0\\
0 & 0& 1& 0& 5& 0 & 5 & 0 &1\\
1 & 0& 0& 0& 0& 5& 0 & 5 &0\\
\end{array} \right) \qquad F^\prime \;=\; {1\over {1380}}\left( \begin{array}{ccccccccc}
0 &5& 5& 5& 5& 45& 45& 91 &  91\\
39 &12& 0& 1& 0& 0& 0& 0 &  0\\
52 &28& 0& 4& 0& 0& 0& 0 &  0\\
39 &0& 12& 0& 1& 0& 0& 0 &  0\\
52 &0& 28& 0& 4& 0& 0& 0 &  0\\
13 &26& 0& 12& 0& 1& 0 & 0& 0\\
32 &20& 0& 28& 0& 4& 0& 0 &  0\\
13 &0& 26& 0& 12& 0& 1& 0 &  0\\
32 &0& 20& 0& 28& 0& 4& 0 &  0\\
1 & 12& 0& 26& 0& 12& 0 & 1 & 0\\
4 &28& 0& 20& 0& 28& 0& 4 &  0\\
1 &12& 0& 26& 0& 12& 0 &  1& 0\\
4 &0& 28& 0& 20& 0& 28& 0 &  4\\
1 &0& 1& 0& 12& 0& 26& 0& 12 \\
4 &0& 4& 0& 28& 0& 20& 0& 28 \\
1 &1& 0& 12& 0& 26& 0& 12 &  0\\
4 &4& 0& 28& 0& 20& 0& 28 &  0\\
1 &0& 1& 0& 12& 0& 26& 0 & 12\\
4 &0& 4& 0& 28& 0& 20& 0 &  28\\
\end{array} \right) \end{equation}
\begin{equation}
C^\prime = {1\over {664}}\left( \begin{array}{ccccccccccccccccc}
0 &27 & 15& 27& 15& 13& 9& 13 & 9 &3 & 3& 3& 3& 3& 3& 3 & 3\\
6 &10 & 6& 0& 0& 14& 6& 0 & 0 &10 & 6& 0& 0& 3& 3& 0 & 0\\
6 &0 & 0& 10& 6& 0& 0& 14 & 6 &0 & 0& 10& 6& 0& 0& 3 & 3\\
6 &3 & 3& 0& 0& 10& 6& 0 & 0 &14 & 6& 0& 0& 10& 6& 0 & 0\\
6 &0 & 0& 3& 3& 0& 0& 10 & 6& 0 &0 & 14& 6& 0& 0& 10& 6\\
22 &0 & 0& 0& 0& 3& 3& 0 & 0 &10 & 6& 0& 0& 14& 6& 0 & 0\\
22 &0 & 0& 0& 0& 0& 0& 3 & 3 &0 & 0& 10& 6& 0& 0& 14& 6 \\
42 &0 & 0& 0& 0& 0& 0& 0 & 0 &3 & 3& 0& 10& 6& 0& 0 & 0\\
42 &0 & 0& 0& 0& 0& 0& 0 & 0 &0 & 0& 3& 3& 0& 0& 10 & 6\\
\end{array} \right)\end{equation}
\begin{equation}
D^\prime = {1\over {4294}}\left( \begin{array}{ccccccccccccccccc}
0 &103 &120 & 28& 36& 28& 36& 4& 8 & 4 &8 & 28& 36& 28& 36& 103& 120 \\
163 &0 & 0& 61& 68& 1& 1& 0 & 0 &13 & 18& 0& 0& 1& 2& 0 & 0\\
60 &0 & 0& 14& 16& 1& 1& 0 & 0 &13 & 14& 0& 0& 1& 2& 0 & 0\\
34 &61 & 68& 0& 0& 61& 68& 0 & 0 &13 & 18& 0& 0& 1& 2& 0 & 0\\
30 &14 & 16& 0& 0& 14& 16& 0 & 0 &13 & 14& 0& 0& 1& 2& 0 & 0\\
34 & 0 & 0 &61 & 68& 0& 0& 61& 68& 0 & 0 &13 & 18& 0& 0& 1& 2\\
30 & 0 & 0 &14 & 16& 0& 0& 14& 16& 0 & 0 &13 & 14& 0& 0& 1& 2\\
6 &13 & 18& 0 & 0 &61 & 68& 0& 0& 61& 68& 0 & 0 &13 & 18& 0& 0\\
6 &13 & 14& 0 & 0 &14 & 16& 0& 0& 14& 16& 0 & 0 &13 & 14& 0& 0\\
6 & 0& 0 &13 & 18& 0 & 0 &61 & 68& 0& 0& 61& 68& 0 & 0 &13 & 18\\
6 & 0& 0&13 & 14& 0 & 0 &14 & 16& 0& 0& 14& 16& 0 & 0 &13 & 14\\
34 & 1 & 2& 0& 0 &13 & 18& 0 & 0 &61 & 68& 0& 0& 61& 68& 0 & 0\\
30 & 1& 2& 0& 0&13 & 14& 0 & 0 &14 & 16& 0& 0& 14& 16& 0 & 0 \\
34 & 0 & 0 & 1 & 2& 0& 0 &13 & 18& 0 & 0 &61 & 68& 0& 0& 61& 68\\
30 & 0 & 0 & 1& 2& 0& 0&13 & 14& 0 & 0 &14 & 16& 0& 0& 14& 16\\
163  &0 & 0&0 & 0 & 1 & 2& 0& 0 &13 & 18& 0 & 0 &61 & 68& 0& 0\\
60  &0 & 0&0 & 0 & 1& 2& 0& 0&13 & 14& 0 & 0 &14 & 16& 0& 0\\
\end{array} \right)\label{Dprimeeq}
\end{equation}
\end{widetext}\normalsize
Note that the matrix is no longer symmetric due to the fact that we are assuming Eve has access only to Bob's line but not Alice's. In particular, this means
that $F^\prime $ is not equal to the transpose of $C^\prime$.

\vfill

\end{document}